\documentclass[journal]{IEEEtran}

\ifCLASSINFOpdf
\else
   \usepackage[dvips]{graphicx}
\fi
\usepackage{url}
\usepackage{subfigure}
\usepackage{pdfpages}
\usepackage{tabularx}

\usepackage{graphicx}

\usepackage[tbtags]{amsmath}

\begin{document}

\title{Are Microphone Signals Alone Sufficient for Self-Positioning? }

\author{Faxian Cao, Yongqiang Cheng, Adil Mehmood Khan, and Zhijing Yang 
\thanks{This work was supported by China Scholarship Council (Corresponding author: Yongqiang Cheng). }
\thanks{F. Cao, Y. Cheng and A. M. Khan are with School of Computer Science, University of Hull, Hull HU6 7RX, U.K. (e-mail: \{faxian.cao-2022, y.cheng, a.m.khan\}@hull.ac.uk).}
\thanks{Z. Yang is with School of Information Engineering, Guangdong University of Technology, Guangzhou 510006, China (e-mail:yzhj@gdut.edu.cn).}}

\markboth{
}
{Shell \MakeLowercase{\textit{et al.}}: Bare Demo of IEEEtran.cls for IEEE Journals}
\maketitle

\begin{abstract}
In an era where asynchronous environments pose challenges to traditional self-positioning methods, we propose a new transformation to the existing paradigm. Traditionally, time of arrival (TOA) measurements require both microphone and source signals, limiting their applicability in environments with unknown emission time of human voices or sources and unknown recording start time of independent microphones. To address this issue, our research pioneers a mapping function capable of transforming both TOA and time difference of arrival (TDOA) formulas, demonstrating, for the first time, that they can be identical to one another. This implies that microphone signals alone are sufficient for self-positioning without the need for source signals waveform, a groundbreaking advancement in the field that carries the potential to revolutionize self-positioning techniques, expanding their applicability in challenging environments. Supported by a robust mathematical proof and compelling experimental results, this research represents a timely and significant contribution to the current discourse in signal, and audio processing.
\end{abstract}


\IEEEpeerreviewmaketitle

\section{Introduction}
\IEEEPARstart{T}{he} ability to accurately localize distributed microphones and sound sources is a fundamental requirement in various acoustic tasks, including noise reduction, source signal enhancement, and separation~\cite{1,2,3}. This is conventionally achieved through the utilization of time of arrival (TOA) and time difference of arrival (TDOA) measurements~\cite{4}. However, these techniques have significant limitations, particularly in asynchronous environments where the timing information of signal emission and recording are unknown in advance. 

In scenario where the waveform of the source signals is available, including information on frequency, amplitude, and duration, TOA measurements can be estimated through cross-correlation methods~\cite{5}. This has led to the development of various self-positioning methodologies such as probabilistic generative models~\cite{6}, maximal likelihood estimation~\cite{7}, Gram matrix and semi-definite relaxation~\cite{8}, and techniques utilizing the low-rank property (LRP)~\cite{9} with alternating minimization method~\cite{10, 11}, and structure total least square~\cite{12, 13}.

Alternatively, when source signals waveform is hard to obtain, self-positioning techniques pivot towards TDOA measurements, which can be estimated with audio signals from a pair of microphones~\cite{5}. This shift has led to a plethora of methodologies, such as maximal likelihood estimation~\cite{7, 18}, auxiliary function method~\cite{14}, LRP with nuclear truncation minimization~\cite{15,16}, and distributed damped Newton optimization~\cite{4, 19}.

Yet, amidst these developments, a significant and pressing question has lingered: Can microphone signals alone be sufficient for self-positioning, thereby negating the need for source signals? The answer to this question carries profound implications for the field, as it can streamline and make the self-positioning process more efficient and adaptable. Moreover, it presents a timely advancement, given the growing complexity of audio environments and the increasing need for flexible and efficient localization methods.

This research takes a groundbreaking step towards answering this crucial question. We introduce an innovative mapping function that transforms both TOA and TDOA formulas to an identical representation/form. In other words, our findings illustrate that their transformations can mirror one another perfectly, confirming that relying solely on microphone signals is sufficient for self-positioning tasks. Therefore, our novel approach unveils, for the first time, the exact relationship between TOA and TDOA measurements, challenging the long-standing assumption that TOA necessitates both microphone signals and the source signal waveform.

This revolutionary insight doesn't merely simplify the self-positioning process by eliminating the need for additional information from source signals. It also broadens its applicability, as properties initially designed for TOA-based localization, such as rank 3~\cite{3} and rank 5~\cite{20}, can now be applied to TDOA-based localization. In essence, our work represents a significant, novel, and timely contribution, with the potential to dramatically reshape self-positioning techniques in asynchronous environments and catalyze further advancements in signal, and audio processing. 
\section{Problem Formulation}
Consider a setup where we have $M$ asynchronous microphones and $N$ asynchronous sound sources, located at $R=\begin{bmatrix}
    r_1,& \cdots, & r_M
\end{bmatrix}_{3 \times M}$ and $S=\begin{bmatrix}
    s_1,& \cdots,& s_N
\end{bmatrix}_{3  \times N}$, respectively, with 3 denoting three dimensions. After sources have emitted the audio signals and microphones have received the corresponding signals, we can encounter two possible scenarios.

In the scenario where the waveform from the source signals can be acquired, by defining the recording start time of $i^{th}$ microphone as $\delta_i$ and  emission time of $j^{th}$ source as $\eta_j$ as well as the speed of sound as $c$, the TOA ($t_{i,j}$) between $i^{th}$ microphone and $j^{th}$ source can be calculated as~\cite{8}
\begin{equation}
\label{eq1}
    t_{i,j}=\frac{\|r_i-s_j\|}{c}+\eta_j-\delta_i,
\end{equation}
where $i=1, \cdots, M$ and $j=1, \cdots, N$, and $\| \bullet \|$ is the $l_2$ norm. In addition, without loss of generality, the location of the first source can be set as $s_1=[0, 0,0]^T$ because of the invariance of translation and rotation regarding the geometry of microphones and sources~\cite{3}.

The second scenario arises when it is challenging to obtain the waveform from the source signals. Here, we define the TDOA ($\tau_{i,j}$) of  $j^{th}$ source between  $i^{th}$ microphone and $1^{st}$ microphone as~\cite{8}
\begin{equation}
\label{eqtd1}
   \tau_{i,j}=t_{i,j}-t_{1,j}=\frac{\|r_i-s_j\|}{c}-\frac{\|r_1-s_j\|}{c}+\delta_1-\delta_i.
\end{equation}

Upon inspection of Eq. (\ref{eqtd1}), it can be observed that after the source $j$ emits the audio signal and both $i^{th}$ and $1^{st}$ microphones receive the corresponding signal, the $i^{th}$ microphone signal contains information about the start time of $i^{th}$ microphone $\delta_i$, the emitted time of  $j^{th}$ source $\eta_j$, as well as the time difference in signal propagation from the $j^{th}$ source to the $i^{th}$ microphone. Similarly, the signal at the $1^{st}$ microphone contains information about the start time of the $1^{st}$ microphone $\delta_1$, the emitted time of  $j^{th}$ source  $\eta_j$, and the time difference in signal propagation from the $j^{th}$ source to the $1^{st}$  microphone. Thus, employing the generalized cross-correlation with phase transform~\cite{21} method, TDOA ($\tau_{i,j}$)  can be estimated using only the audio signals from the $j^{th}$ and $1^{st}$ microphones, demonstrating the independence of TDOA from the source signal. Besides, according to the definition of TDOA, it measures the time difference between a pair of microphones when they receive the corresponding source signal, therefore, the TDOA ($\tau_{i,j}$) of $j^{th}$ source in Eq. (\ref{eqtd1}) can also be measured by the $i^{th}$ microphone signal and any other one of remaining microphone signals. 

Let’s denote $\Check{\delta}_i=\delta_i-\delta_1$ and $\Check{\eta}_j=-\frac{\|r_i-s_j\|}{c}$, then TDOA formula in Eq. (\ref{eqtd1}) can be re-written as~\cite{8}
\begin{equation}
\label{pf1}
   \tau_{i,j}=\frac{\|r_i-s_j\|}{c}+\check{\eta}_j-\check{\delta}_i.
\end{equation}

Interestingly, this equation shares the same structural form as the TOA formula in Eq. (\ref{eq1}). However, the exact relationships between TOA formula in Eq. (\ref{eq1}) and TDOA formula in Eq. (\ref{eqtd1}) remain elusive. No existing works have demonstrated this relationship so far, and as a result, the sufficiency of utilizing only microphone signals for self-positioning is still unknown. Our research objective, therefore, is to investigate the feasibility of utilizing the microphone signals alone for self-positioning when the waveform of source signals is unavailable. The results of our study have the potential to challenge the long-standing assumption that the acquisition of source signal waveform is a necessity for TOA-based self-positioning. This can lead to an expansion of self-positioning techniques, enhancing their utility in challenging environments.

\section{Mapping function for TOA and TDOA Formulas}
\label{sec:guidelines}

In this section, a novel mapping function is derived for TOA formula in Eq. (\ref{eq1}) and TDOA formula in Eq. (\ref{eqtd1}). We first present the novel mapping function in Subsection A, then the proof of the proposed mapping function is shown in Subsection B followed by a subsection for showing the property of the proposed mapping function. 
\subsection{Mapping Function $f(\bullet)$}
TOA measurements are unavailable when waveform of source signals is missing, and only TDOA measurements can be used for localization once this situation happens. Since there are no existing works in the state-of-the-arts investigate the relationships between TOA and TDOA measurements, here, we present a novel mapping function to show the sufficiency of using microphone signals alone for both TOA and TDOA-based self-positioning. The proposed mapping function, $f(\bullet)$, for TOA formula in Eq. (\ref{eq1}) and TDOA formula in Eq. (\ref{eqtd1}) is defined as
\begin{equation}
    \label{cf}
    f(t_{i,j})=t_{i,j}-t_{i,1}-\frac{\sum_{i=1}^{M}(t_{i,j}-t_{i,1})}{M},
\end{equation}
and
\begin{equation}
    \label{cf1}
    f(\tau_{i,j})=\tau_{i,j}-\tau_{i,1}-\frac{\sum_{i=1}^{M}(\tau_{i,j}-\tau_{i,1})}{M},
\end{equation}
respectively, then by applying the mapping function, $f(\bullet)$, to TOA formula in Eq. (\ref{eq1}) and TDOA formula in Eq. (\ref{eqtd1}) and defining two variables
\begin{equation}
    \label{add1}
    \begin{cases}
      \dot{\delta}_i=\frac{\|r_i-s_1\|}{c} \\
      \dot{\eta}_j=\frac{\sum_{i=1}^{M}(\|r_i-s_1\|-\|r_i-s_j\|)}{cM}
    \end{cases},
\end{equation}
we state that\vspace{-1em}
 \begin{equation}
             \label{cf2}
        f(t_{i,j})= f(\tau_{i,j}) = \frac{\|r_i-s_j\|}{c}- \dot{\delta}_i+\dot{\eta}_j,
 \end{equation}
where $i=1, \cdots, M$ and $j=1, \cdots, N$. From the statement in Eq. (\ref{cf2}), we can see that once this relationship is proved, this mapping function indicates the same structure as TOA formula in Eq. (\ref{eq1}), showing the location of both microphones and sources can be obtained with $f(\bullet)$ by utilizing the same methods that are designed for TOA-based self-positioning. More importantly, the sufficiency that utilizing microphone signals alone can be revealed for self-positioning, providing the potential to challenge the long-standing assumption that TOA necessitates both microphone signals and the waveform of source signals for self-positioning. Thus, the process of self-positioning can be more adaptable and efficient, and the abilities of self-positioning techniques can be expanded in challenging environments. 
\subsection{Proof for mapping function $f(\bullet)$}
We first derive the transformation of TOA formula in Eq. (\ref{cf}), then the derivation of transformation of TDOA formula in Eq. (\ref{cf1}) is displayed. Finally, we validate the statement in Eq. (\ref{cf2}) by comparing the transformation of TOA formula in Eq. (\ref{cf}) with the transformation of TDOA formula in Eq. (\ref{cf1}).

\subsubsection{Transformation of TOA formula}
{From TOA formula in Eq. (1), we can have
\begin{equation}
    \label{eq2}
     t_{i,1}=\frac{\|r_i-s_1\|}{c}+\eta_1-\delta_i,
\end{equation}
then with Eqs. (\ref{eq1}) and (\ref{eq2}), the difference between $t_{i,j}$ and $t_{i,1}$ can be displayed as
\begin{equation}
 \label{eq3}
     t_{i,j}-t_{i,1}=\frac{\|r_i-s_j\|}{c}-\frac{\|r_i-s_1\|}{c}+\eta_j-\eta_1.
\end{equation}

From Eq. (\ref{eq3}), we can see the mean value for $t_{i,j}-t_{i,1}$ with respect to the index $i$ is
\begin{equation}
\label{peq6}
    \frac{\sum_{i=1}^{M} (t_{i,j}-t_{i,1})}{M}=\frac{\sum_{i=1}^M(\|r_i-s_j\|-\|r_i-s_1\|)}{cM}+\eta_j-\eta_1.
\end{equation}

Finally, with Eqs. (\ref{eq3}) and (\ref{peq6}), the mapping function for TOA formula in Eq. (\ref{cf}), $f(t_{i,j})$, can be formulated as
\begin{small}
\begin{align}
   \label{cf8}
      &f(t_{i,j})=t_{i,j}-t_{i,1}-\frac{\sum_{i=1}^{M}(t_{i,j}-t_{i,1})}{M} \nonumber \\
     & =\frac{\|r_i-s_j\|}{c}-\frac{\|r_i-s_1\|}{c}+\frac{\sum_{i=1}^M(\|r_i-s_1\|-\|r_i-s_j\|)}{cM},
\end{align}
\end{small}
where $i=1, \cdots, M$ and $j=1, \cdots, N$.
}

\subsubsection{Transformation of TDOA formula} {From TDOA formula in Eq. (\ref{eqtd1}), we can display the difference between $\tau_{i,j}$ and $\tau_{i,1}$ as
\begin{align}
     \label{eqtd5}
     \tau_{i,j}-\tau_{i,1}=& \frac{\|r_i-s_j\|}{c}-\frac{\|r_1-s_j\|}{c} \nonumber \\
     &-\frac{\|r_i-s_1\|}{c}+\frac{\|r_1-s_1\|}{c},
\end{align}
then from Eq. (\ref{eqtd5}), the mean value for $\tau_{i,j}-\tau_{i,1}$ with respect to the index $i$ can be written as
\begin{align}
    \label{eqtd7}
   \frac{\sum_{i=1}^{M}(\tau_{i,j}-\tau_{i,1})}{M}=&\frac{\sum_{i=1}^M(\|r_i-s_j\|-\|r_i-s_1\|)}{cM} \nonumber \\
     & -\frac{\|r_1-s_j\|}{c}+\frac{\|r_1-s_1\|}{c}.
\end{align}

Finally, with Eqs. (\ref{eqtd5}) and (\ref{eqtd7}), the mapping function for TDOA formula in Eq. (\ref{cf1}), $f(\tau_{i,j})$, can be formulated as
\begin{small}
\begin{align}
   \label{cf10}
    & f(\tau_{i,j})=\tau_{i,j}-\tau_{i,1}-\frac{\sum_{i=1}^{M}(\tau_{i,j}-\tau_{i,1})}{M} \nonumber \\
     &=\frac{\|r_i-s_j\|}{c}-\frac{\|r_i-s_1\|}{c}+ \sum_{i=1}^M\frac{(\|r_i-s_1\|-\|r_i-s_j\|)}{cM}
\end{align}
\end{small}
 where $i=1, \cdots, M$ and $j=1, \cdots, N$. 
}

\subsubsection{Validation of statement}
{Based on the definitions of the two variables $\dot{\delta}_i$ and $\dot{\eta}_j$ in Eq. (\ref{add1}), then with the transformation of TOA formula in Eq. (\ref{cf8}) and transformations of TDOA formula in Eq. (\ref{cf10}), we can see that Eq. (\ref{cf8}) and  Eq. (\ref{cf10}) are identical to one another, this completes the proof of mapping function $f(\bullet)$ in Eq. (\ref{cf2}). 

With the proof of the statement in Eq. (\ref{cf2}), we can see that the transformations of TOA and TDOA formulas are identical to one another, revealing the sufficiency of utilizing microphone signals for both TOA and TDOA-based self-positioning, providing the potentials to challenge the long-standing assumption that TOA necessitates both microphone signals and the waveform of source signals for self-positioning. In addition, the statement in Eq. (\ref{cf2}) indicates that many properties, such as rank 3~\cite{3} and rank 5~\cite{20}, that are used for TOA-based localization can also be used for TDOA-based localization, this makes the tasks of self-positioning more efficient and adaptable. Besides, by eliminating the need for additional information from source signals, a wide range of other applications, such as noise reduction, sources signals enhancement and separation~\cite{1, 2, 3} can also be facilitated since the importance of self-positioning for those applications above.
} 
\subsection{Property for Mapping Function $f(\bullet)$}\label{formats}

Since $s_1=[0,   0,0]^T$ (see content below Eq. (\ref{eq1})), let’s denote $x_i=\frac{\|r_i\|}{c}$ and $y_{i,j}=\frac{\|r_i-s_j\|}{c}$ for $i=1,\cdots, M$ and $j=1, \cdots, N$. Then based on Eqs. (\ref{cf2}), (\ref{cf8}) and (\ref{cf10}), we can summarize the proposed mapping function, $f(\bullet)$, as close form
\begin{equation}
\label{add4}
    f(\bullet)=(y_{i,j}-\frac{\sum_{i=1}^{M}y_{i,j}}{M})-(x_{i}-\frac{\sum_{i=1}^{M}x_{i}}{M}).
\end{equation}

Finally, with the close form of proposed mapping function in Eq. (\ref{add4}), we can see the interesting property of this mapping function, i.e., the mean value of mapping function $f(\bullet)$ is 0 with respect to index $i$, which can be summarized as
\begin{equation}
    \label{add5}
    \frac{\sum_{i=1}^{M}f(\bullet)}{M}=    \frac{\sum_{i=1}^{M}f(t_{i,j})}{M}=\frac{\sum_{i=1}^{M}f(\tau_{i,j})}{M}=0.
\end{equation}

Upon inspection of Eq. (\ref{add5}), it indicates that the mean value of transformation of both TOA and TDOA formulas with respect to all microphones and any source is 0. 
\section{Experimental Validations}
In this section, experimental results are shown to validate the proposed mapping function. The experimental setups are illustrated in subsection A first, then the evaluation metric is defined and the validations of both the proposed mapping function and the property of the proposed mapping function are shown in subsection B. 
\subsection{Setups}
\subsubsection{Simulation data}{All the simulation data is randomly generated by MATLAB with uniform distribution, both the start time of microphones and emission time of sources are in the range of $[-1, \ 1]s$, the locations of microphone and source are distributed in the room with size of $10 \times 10 \times 3$ $m^3$~\cite{8} and the speed of sound is set to be $340m/s$. In addition, both the number of microphones $M$ and the number of sources $N$ are set to $20$, and the number of configurations is set to be $1000$. Besides, since the number of both microphones and  sources is $20$ and the number of configurations is $1000$, there are $400000$ data points for simulated data.} 
\subsubsection{Real-Life data}{The real data~\cite{22} was collected in an office of size of  $5 \times 3$ $m^2$,  where most of the furniture inside the office was removed. There are $12$ microphones which were fixed, and a chirp was played by a loudspeaker from $65$ positions. This real-life data for $12 \times 65$ TOA matrix can be downloaded at Github\footnote{This real-life data is available at \url{https://github.com/swing-research/xtdoa/tree/master/matlab}}~\cite{8, 22} and the TDOA matrix is calculated by Eq. (\ref{eqtd1}). For more details of this real-life data, readers can refer to references~\cite{8, 22}. Also, both the start time of microphones and the emission time of sources are in the range of $[-1,\ 1]s$.  In addition, the number of data points for real-life data is 780 since there are 12 microphones and 65 sources.} 
\subsection{Evaluations and Results}
We first show the value of proposed mapping function for the transformations of both TOA and TDOA measurements with both simulation data and real-life dataset, then the property of the proposed mapping function in Eq. (\ref{add5}),  $\frac{\sum_{i=1}^{M}f(\bullet)}{M}=0$, is validated. Finally, the statement for the proposed mapping function in Eq. (\ref{cf2}) is evaluated by measuring the difference of transformations of TOA and TDOA formulas
\begin{equation}
\label{eval}
    \Delta f_{i,j}=f(t_{i,j})-f(\tau_{i,j}),
\end{equation}
where $i=1,\cdots, M$ and $j=1,\cdots,N$. As can be seen from Eq. (\ref{eval}), once $\Delta f_{i,j}$ is equal to zero, the values of transformation of TOA and TDOA formulas are the same as each other, it indicates that the proposed mapping function, $f(\bullet)$, is validated. 

\begin{figure}[t]
\centerline{\subfigure[]{\includegraphics[trim=0.2cm 0.2cm 1.3cm 0.2cm, clip,
width=0.48\columnwidth]{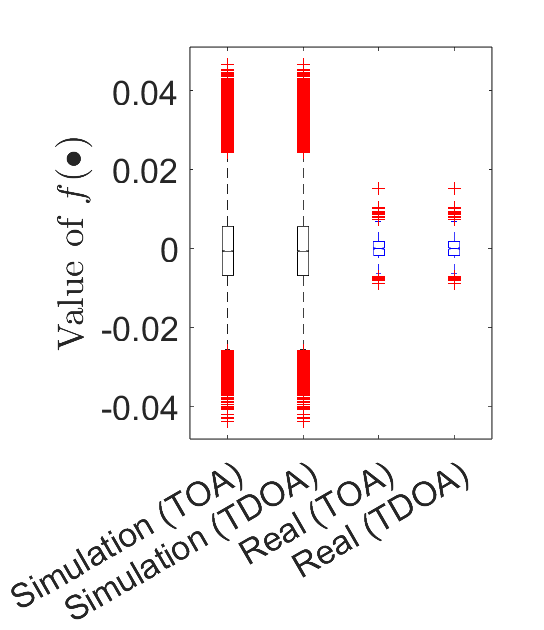}}
\subfigure[]{\includegraphics[trim=0.2cm 0.2cm 1.3cm 0.2cm, clip,
width=0.48\columnwidth]{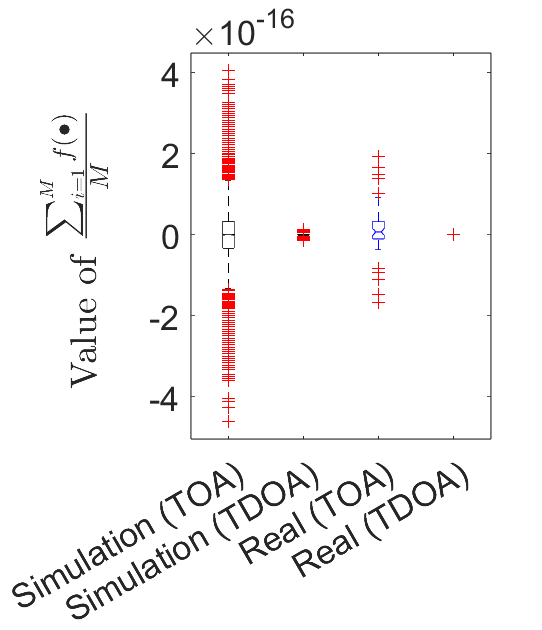}}}
\subfigure[]{\centerline{\includegraphics[trim=0.2cm 0.2cm 1.3cm 0.2cm, clip,
width=0.48\columnwidth]{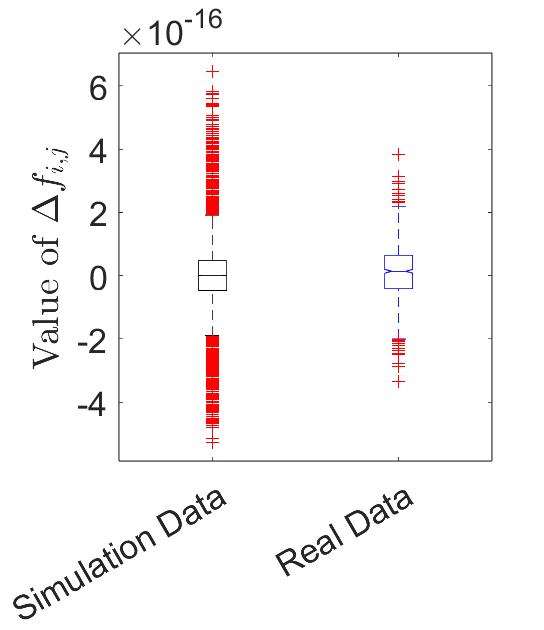}}}
\caption{Validation of proposed mapping function with both simulation and real-life data sets: (a) the value of proposed mapping function $f(\bullet)$ for  transformations of both TOA measurements in Eq. (\ref{cf}) and TDOA measurements in Eq. (\ref{cf1}); (b) the value of  $\frac{\sum_{i=1}^{M}f(\bullet)}{M}$ in Eq. (\ref{add5}) for transformations of both TOA and TDOA measurements; (c) the value of  $\Delta f_{i,j}$ in Eq. (\ref{eval}).}
\label{results}
\end{figure}

Fig. \ref{results} shows the experimental results with both simulated data and real data. From Fig. \ref{results}(a), it can be observed that the values of $f(\bullet)$ for both TOA and TDOA measurements in simulated data are in the range of $[-0.05, \ 0.05]$ while the corresponding values in real data are in the range of $[-0.02, \ 0.02]$, this is because of the different sizes of the rooms are used for simulation data and real life data, respectively. In addition, form Fig. \ref{results}(b), we can see that the values of $\frac{\sum_{i=1}^{M}f(\bullet)}{M}$ in both simulation and real data are always with a magnitude of $10^{-16}$, and it should be noted that those errors/inaccuracies are introduced by the machine calculation accuracy of MATLAB. Therefore, the property of proposed mapping function $\frac{\sum_{i=1}^{M}f(\bullet)}{M}=0$ is validated. Besides, from Fig. \ref{results}(c), we can also see that the value of $\Delta f_{i,j}$ is also with a magnitude of $10^{-16}$ due to the machine calculation accuracy of MATLAB, therefore, $\Delta f_{i,j}=0$ is validated. This implies that the transformations of TOA formula in Eq. (\ref{eq1}) and TDOA formula in Eq. (\ref{eqtd1}) are identical to one another, so that the statement for proposed mapping function in Eq. (\ref{cf2}) is validated. TOA measurements are obtained with both microphones received signals and source signals while TDOA measurements are obtained with microphones signals only, and from Fig. \ref{results}(c), it is obvious that the transformation of TOA and TDOA measurements are the same as each other, therefore, our novel mapping function shows the sufficiency of utilizing microphone signals alone for self-positioning, negating the need of source signals for self-positioning, presenting a timely advancement for tasks of self-positioning.
\section{Conclusion}
This letter investigated the sufficiency of using microphone signals alone for self-positioning that has never been investigated in the state-of-the-arts. When both the emission time of the source signal and the recording start times of the microphones are unknown, by presenting a novel mapping function that has never been shown in the literature to transform both TOA and TDOA formulas, we demonstrated that the transformations of TOA and TDOA formulas are identical to one another, showing the sufficiency that uses microphone signals alone for self-positioning, making the tasks of self-positioning more flexible and adaptable. Therefore, the proposed mapping function can be regarded as a timely advancement for the tasks of self-positioning.

For future works, based on the existing TOA and TDOA-based methods, it would be interesting to apply this mapping function to estimate the unknown emission time and start time as well as the locations of microphones and sources. Besides, it might also be interesting to apply this mapping function to other applications, such as noise reduction, sources signals enhancement and separation.

\end{document}